# Magnetoresistivity in MgB$_2$ as a probe of disorder in π- and σ-bands

I.Pallecchi[1], V.Ferrando[1,4], E.Galleani D'Agliano[1], D.Marré[1], M.Monni[1], M.Putti[1], C.Tarantini[1], F.Gatti[2], H.U.Aebersold[3], E.Lehmann[3], X.X.Xi[4], E.G.Haanappel[5], C.Ferdeghini[1]

[1] LAMIA-CNR- INFM and Università di Genova, via Dodecaneso 33, 16146 Genova Italy

[2] Università di Genova, via Dodecaneso 33, 16146 Genova Italy

[3] Paul Scherrer Institut, CH-5232 Villigen, Switzerland

[4] The Pennsylvania State University, University Park, PA 16802, USA

[5] Laboratoire National des Champs Magnétiques Pulsés, CNRS-UPS-INSA, Toulouse, France

**Abstract**

In this paper we present normal state magnetoresistivity data of magnesium diboride epitaxial thin films with different levels of disorder, measured at 42K in magnetic fields up to 45 Tesla. Disorder was introduced in a controlled way either by means of neutron irradiation or by carbon doping. From a quantitative analysis of the magnetoresistivity curves with the magnetic field either parallel or perpendicular to the plane of the film, we extract the ratio of the scattering times in π- and σ- bands. We demonstrate that the undoped unirradiated thin film has π scattering times smaller than σ ones; upon irradiation, both bands become increasingly more disordered; eventually the highly irradiated sample (neutron fluence~$7.7 \cdot 10^{17}$ cm$^{-2}$) and the C-doped sample have comparable scattering times in the two types of bands. This description of the effect of disorder in the two kinds of bands on transport is consistent with the residual resistivity values and with the temperature dependence of the resistivity.

**I. Introduction**

The superconducting and normal state properties of magnesium diboride[1] are strongly influenced by the presence of two types of bands of different character[2,3] at the Fermi level. Four bands in total cross the Fermi level of $MgB_2$. Two of these bands, usually called π-bands, are formed by the $p_z$ orbitals of boron atoms; they are weakly coupled to the phonons, they have three-dimensional character, one of them has electron-like charge carriers and the other one has hole-like ones. The other two bands, the σ-bands, are formed by $sp^2$-hybridized orbitals stretched along boron-boron bonds and are two-dimensional, hole-type and strongly coupled with the optical $E_{2g}$ phonon mode. The different parity of π- and σ- bands inhibits the interband scattering, turning these two sets of bands into parallel, almost independent conducting channels. The ratio of the scattering times in the π- and σ-bands is a crucial parameter in determining normal state and superconducting transport properties[4]. Indeed, by selective doping in the two bands it is possible to drive them from the clean into the dirty regime, thus tailoring many properties such as infrared reflectivity[5], microwave conductivity[6] and upper critical field[7]. In particular, the upper critical field $B_{c2}$ depends on the ratio of the intraband electronic scattering times in the two sets of bands, as predicted by Gurevich[8] and by Golubov and Koshelev[9]. Because the transport properties of $MgB_2$ are determined by two conducting channels in parallel, the zero-temperature value of $B_{c2}(0)$ depends only on the largest scattering time, whereas the normal-state resistivity is governed by both scattering times. Therefore, it is possible that samples with resistivity values which vary by more than one order of magnitude exhibit comparable critical fields[10].

Concerning the electrical resistivity ρ, its values just before the superconducting transition may vary in a wide range, typically from a few tenth to a few hundreds μΩ·cm in different samples. As such values of ρ are determined both by intrinsic and intergrain contributions, in samples with poor intergrain connectivity the ratio of the scattering times in the two types of bands cannot be extracted within a multi-band model[11]. Instead, the temperature dependence of the resistivity is less sensitive to intergrain connectivity and is therefore a better parameter to describe the intrinsic transport and also to get information on the contribution of the two types of bands. In this context, the effect on the critical temperature and resistivity of introducing a controlled amount of disorder by irradiation with alpha particles has been studied by Gandikota et al.[12], while Bugoslavsky et al.[13,] have extracted the scattering rates from the fitting of ρ(T) of disordered thin films.

Magnetoresistivity data above $T_c$ are even more sensitive to the multiband character of transport in magnesium diboride and they vary within a broad range of values[14,15,16,17,18]. Large positive magnetoresistivity is usually found in clean $MgB_2$ samples. At low magnetic fields, magnetoresistivity is proportional to the square of the applied field, but at larger fields - namely when the cyclotron frequency $\omega_c = e \cdot B/m_{eff}$ becomes larger than the scattering rate - it tends to saturate. In oriented thin films, the scenario is complicated by the anisotropy of the band structure. In order to extract quantitative information from magnetoresistivity data in both configurations, i.e. with B parallel and perpendicular to the boron planes (*ab* planes in the following), a picture is necessary which takes into account the presence of all four bands and terms of order higher than $B^2$ in the magnetic field in the expression for the magnetoresistivity.

In this paper we present the expression for magnetoresistivity resulting from such a picture and we apply it to the analysis of magnetoresistivity data up to 45 Tesla, measured in oriented thin films where disorder is introduced in a controlled way, either by neutron irradiation or by carbon doping. Indeed, a systematic study of the effect of disorder needs on the one hand a series of samples with a well-defined kind of disorder which steadily increases from sample to sample and on the other hand measurements carried out up to very high magnetic fields, so that even in the most disordered sample an appreciable magnetoresistivity is still detectable.

In section II the calculation of the magnetoresistivity is described. In section III, the sample preparation and the measurement technique are explained. Then, the magnetoresistivity data and the results of the fits are presented, discussed and compared with the measured values of the resistivity and its temperature derivative. Finally, the conclusions of the work are summarized.

## II. Calculation of anisotropic multi-band magnetoresistivity

According to the theory of cyclotron orbits, positive transverse (that is with B perpendicular to the current) magnetoresistivity at small fields is expected to have a leading term proportional to the square of the magnetic field. In the case of a single, spherically symmetric band, the standard Boltzmann equation approach gives a zero transverse magneto-resistance[19]. Non-vanishing magnetoresistivity can be due either to a non-spherical Fermi surface or to an anisotropic $\tau(\underline{k})$ scattering time[20]; however, the generalization of the theory to the case of many uncoupled bands also leads to a non-vanishing magnetoresistivity, unless the bands are identical. Indeed, the contributions of different bands either cancel or add to each other depending on whether the charge carriers are of the same sign or of opposite sign, respectively. The general formula for the magnetoresistivity, $\frac{\Delta\rho}{\rho(0)} = \frac{\rho(B) - \rho(0)}{\rho(0)}$, in the case of four bands, assuming a diagonal mass tensor with different in-plane and out-of-plane effective masses within the free electron approximation can be extracted starting from the Boltzmann equation. When the magnetic field is perpendicular to the *ab* planes, the magnetoresistivity expression is the same as for the isotropic case, except for the fact that mobilities and conductivities are all calculated along the *ab* planes. Thereby, in the experimental configuration $B \perp ab$, the magnetoresistivity is given by the same expression derived in reference for the isotropic case. This expression, which in equation 3 of ref. 18 was written only to the lowest order in $B^2$, becomes, to all orders in $B^2$:

$$\left.\frac{\Delta\rho}{\rho(0)}\right|_{B\perp ab} = \frac{\sum_{\substack{i,j,k,l \\ i<j<k<l}} \sigma_{(ab)i}\sigma_{(ab)j}\left(1+\mu_{(ab)k}^2 B^2\right)\left(1+\mu_{(ab)l}^2 B^2\right)\left(\mu_{(ab)i}-\mu_{(ab)j}\right)^2 B^2}{\left[\sum_i\left(\sigma_{(ab)i}^2 \prod_{j\neq i}\left(1+\mu_{(ab)j}^2 B^2\right)\right)\right] + 2\sum_{\substack{i,j \\ j>i}}\left(\sigma_{(ab)i}\sigma_{(ab)j}\left(1+\mu_{(ab)i}\mu_{(ab)j}B^2\right)\prod_{\substack{k,l \\ l>k \\ l\neq i, l\neq j, k\neq i, k\neq j}}\left(1+\mu_{(ab)k}^2 B^2\right)\left(1+\mu_{(ab)l}^2 B^2\right)\right)} \quad (1a)$$

On the other hand, when the magnetic field is parallel to the *ab* planes, the expression for magnetoresistivity is even complicated by the fact that cyclotron orbits are not confined only within *ab* planes, but are perpendicular to them; in this case we derive the following expression:

$$\left.\frac{\Delta\rho}{\rho(0)}\right|_{B\|ab} = \frac{\sum_{\substack{i,j \\ i<j}}\left[\left(\sigma_{(c)i}\sigma_{(ab)j}\mu_{(c)j} - \sigma_{(c)j}\sigma_{(ab)i}\mu_{(c)i}\right)\left(1+\mu_{(c)k}\mu_{(ab)k}B^2\right)^2\left(1+\mu_{(c)l}\mu_{(ab)l}B^2\right)\left(\mu_{(ab)j} - \mu_{(ab)i}\right)B^2 \prod_{\substack{k,l \\ l>k \\ l\neq i,l\neq j,k\neq i,k\neq j}}\left(\left(1+\mu_{(c)k}\mu_{(ab)k}B^2\right)^2\left(1+\mu_{(c)l}\mu_{(ab)l}B^2\right)^2\right)\right]}{\left[\sum_i\left(\sigma_{(ab)i}\sigma_{(c)i}\left(1+\mu_{(c)i}\mu_{(ab)i}B^2\right)\prod_{j\neq i}\left(1+\mu_{(c)j}\mu_{(ab)j}B^2\right)^2\right)\right] + \left[\sum_{\substack{i,j \\ j>i}}\left(\left(\sigma_{(c)i}\sigma_{(ab)j}\left(1+\mu_{(c)j}\mu_{(ab)i}B^2\right) + \sigma_{(c)j}\sigma_{(ab)i}\left(1+\mu_{(c)i}\mu_{(ab)j}B^2\right)\right)\left(1+\mu_{(c)i}\mu_{(ab)i}B^2\right)\left(1+\mu_{(c)j}\mu_{(ab)j}B^2\right) \prod_{\substack{k,l \\ l>k \\ l\neq i,l\neq j,k\neq i,k\neq j}}\left(1+\mu_{(c)k}\mu_{(ab)k}B^2\right)^2\left(1+\mu_{(c)l}\mu_{(ab)l}B^2\right)^2\right)\right]}$$

(1b)

In the above equations the subscripts *(ab)* and *(c)* indicate the spatial direction parallel to the *ab* planes or to the *c* axis. The mobility and conductivity of the $i^{th}$-band are related to each other through the carrier concentrations per unit cell in the $i^{th}$-band $n_i$: $\mu_i = V\sigma_i/(n_i e)$, where *V* is the unit cell volume and *e* the carrier charge *including sign* (positive for holes and negative for electrons). The indeces of the sums and products in equations (1) run over all four bands $\pi_1$, $\pi_2$, $\sigma_1$, $\sigma_2$. From equations (1), due to the presence of the linear or squared factor $(\mu_{(ab)j}-\mu_{(ab)i})$ in the numerator, it turns out that the magnetoresistivity is dramatically sensitive to the fact that one of the π-bands, namely $\pi_1$, is electron-like, whereas the other π-band as well as both σ-bands are hole-like. Moreover, the different effective masses of each band in the in-plane and out-of-plane directions also contribute to determine the shape of the magnetoresistivity curves. Of course, theoretical estimates of the effective masses and charge concentrations have to be used in order to fix some of the parameters that enter into our model. In particular, we take for our calculations the values given in table I which have been calculated on the basis of the electronic band structure calculations of Profeta *et al.* [21].

In addition, for simplicity, we assume an average scattering time $\tau_\pi$ in $\pi_1$- and $\pi_2$-bands and similarly an average scattering time $\tau_\sigma$ in both σ-bands. With these assumptions, all the mobility values scale as the inverse ratio of the respective masses multiplied by the ratio of the respective scattering times. Thereby, the magnetoresistivity curves for *B* parallel and perpendicular to the *ab* planes can be obtained from only two free parameters, namely the ratio of the scattering times

$b=\tau_\pi/\tau_\sigma$ and a scaling factor $a$ common to all the mobilities. In practice, the parameter $a$ is chosen to be equal to the absolute value of the $\pi_1$-band mobility along the $c$ axis $|\mu_{(c)\pi 1}|$ in units $m^2 \cdot V^{-1} \cdot s^{-1}$; with this choice $\mu_{(c)\pi 1}=-a$ and any other mobility $\mu_{(dir)\delta}$ is obtained from $\mu_{(c)\pi 1}$ as:

$$\mu_{(dir)\delta} = \pm \frac{|\mu_{(c)\pi 1}| \cdot m_{eff(c)\pi 1}}{\tau_\pi} \frac{\tau_\delta}{m_{eff(dir)\delta}} \quad dir = (ab)\ or\ (c);\quad \delta = \sigma 1, \sigma 2\ or\ \pi 2$$

The fitting parameters also determine the calculated in-plane resistivity of the sample $\rho_{in\text{-}plane}^{(calc)}$, which must be compared with the experimental value at zero field:

$$\rho_{in\text{-}plane}^{(calc)} = \left( \frac{e \cdot n_{\pi 1} \cdot |\mu_{(ab)\pi 1}|}{V} + \frac{e \cdot n_{\pi 2} \cdot \mu_{(ab)\pi 2}}{V} + \frac{e \cdot n_{\sigma 1} \cdot \mu_{(ab)\sigma 1}}{V} + \frac{e \cdot n_{\sigma 2} \cdot \mu_{(ab)\sigma 2}}{V} \right)^{-1} \quad (2)$$

In figure 1, calculated magnetoresistivity curves are presented for three different values of the parameter $b$ and for corresponding values of the parameter $a$ chosen in such a way that the in-plane resistivity is fixed in all cases to $\rho = 5\ \mu\Omega\cdot cm$, a value representative of those of the samples presented here. It can be seen that there is a crossover: the magnetoresistivity in the configuration with $B$ perpendicular to the $ab$ planes is larger than in the configuration with $B$ parallel to the $ab$ planes for small $b$ values and becomes smaller with increasing $b$, that means as σ-bands become dirtier relative to π-bands. The crossover value of $b$ for which the curves in the two configurations nearly overlap depends on the value of the other parameter $a$ and it is $b\sim 1.70$ for $a$ values smaller than 0.02. In the configuration with $B$ parallel to the $ab$ planes, the magnetoresistivity is dominated by the π-bands contribution, due to the very large effective mass of σ carriers in the out-of-plane direction. On the other hand, the magnetoresistivity in the configuration with $B$ perpendicular to the $ab$ planes involves in-plane cyclotron orbits and is sensitive to all four bands; as the parameter $b$ decreases, the π-bands become comparatively dirtier than the σ-bands and the corresponding magnetoresistivity increases. This behavior can be understood by considering that at low $b$ values the $B\perp ab$ magnetoresistivity is large because it is mainly due to the additive contribution of only two bands of opposite sign carriers; on the other hand, at large $b$ values the $B\perp ab$ magnetoresistivity decreases because two more bands with charge carriers of the same sign contribute, providing more cancelling terms proportional to $(\mu_{(ab)j}-\mu_{(ab)i})$ in equations (1).

### III. Experimental procedure

The samples presented in this work were grown on 4H-SiC by Hybrid Physical Chemical Vapor Deposition (HPCVD), using the standard procedure described in detail in ref. [22]. This technique produces high quality epitaxial magnesium diboride films, with low residual resistivity $\rho_0$ (0.3-4 $\mu\Omega$cm) and high critical temperature (~41K) [23]. The main properties of the films are summarized in table II. One of the measured samples, referred to as TS, is a nominally pure, but extremely thin (800-1000 Å) film, so that its resistivity and $T_c$ are respectively higher and lower than the typical values reported for thicker samples [23]; also its residual resistivity ratio RRR is about 4, which is fairly low compared to standard samples grown by HPCVD. A series of films (IRR10, IRR20, IRR30 and IRR40) ), all of them about 2000 Å thick and deposited simultaneously, was neutron irradiated at the spallation neutron source SINQ (thermal neutron flux density up to $1.6 \cdot 10^{13}$ cm$^{-2}$s$^{-1}$) of the Paul Scherrer Institute (PSI) in Villigen, with increasing neutron fluence up to $10^{19}$ cm$^{-2}$. Irradiation by thermal neutrons is an effective way to introduce controlled disorder in MgB$_2$ [24,25,26]. The created defects are mainly due to neutron capture by $^{10}$B, followed by the emission of an alpha particle and a $^7$Li nucleus. Due to the large cross section of this reaction, in samples produced with natural boron (80% of $^{11}$B and 20% of $^{10}$B) the penetration depth of neutrons is as low as 200 µm; however, the thickness of our films is much lower than this penetration depth, so that the irradiation-induced damage is expected to be homogeneous. Indeed, with increasing neutron fluence the residual resistivity $\rho_0$ increases from 1 to 69 µΩ·cm, the critical temperature is suppressed from 41K down to 17K and the residual resistivity ratio RRR passes from 9.8 to almost 1 in the most strongly irradiated film, IRR40 (see table II). As a confirmation of the homogeneous defects distribution the transition width remains sharp, except for the most irradiated sample IRR40. The least irradiated sample IRR10, with a residual resistivity as low as 1 µΩ·cm and a critical temperature as high as 41K, has similar properties as the unirradiated samples of the same thickness. A complete analysis of the series of neutron irradiated films will be reported in a future publication. The last measured sample (referred to as C-dop in the following) has been doped with carbon by adding bis(methylcyclopentadyenil)magnesium ( $(MeCp)_2Mg$ ) to the gas flowing into the HPCVD reactor; the rate of secondary hydrogen flow passing through the $(MeCp)_2Mg$ bubbler determines the overall carbon content in the film[27], which is in our case 7.5% boron atomic concentration. Since $T_c$ has decreased by only 2K with respect to the undoped samples, remaining as high as 39K, it is likely that only a fraction of the carbon content of the film actually substitutes boron, while the remaining carbon is located at the grain boundaries. Comparing the variation of $T_c$ with those observed in C-doped single crystals, we can estimate an actual doping level of less than 2% for this film[28]. Indeed, the contribution of the grain boundaries to the resistivity probably plays a role in increasing the intrinsic value to the measured value of 20 µΩ·cm. Also the RRR value of 2.3 is high compared with the values of samples with similar resistivity, confirming the extrinsic character of the measured resistivity.

High field magnetoresistance measurements were carried out at the Laboratoire National des Champs Magnétiques Pulsés (LNCMP) in Toulouse. For this experiment we have used a 60 T/3 MJ pulsed magnet (rise time to maximum field 50 ms, total pulse duration ~300 ms). The resistance was measured using a standard AC technique with a lock-in frequency between 2 kHz and 20 kHz; for the less resistive samples, the frequency was kept as low as possible in order to minimize the out-of-phase component of the signal. The measurement current was 500 µA for all samples. A rotating sample holder allowed us to change the field direction *in situ*. Measurements were done for a number of temperatures between 4.2K and 42K. In this paper we discuss only the normal state data at 42K. These data were measured up to about 45T, because higher fields were not needed for the analysis and because the higher field results were somewhat noisier due to the field's larger ramping rate.

## IV. Results and discussion

We have measured normal state magnetoresistivity of five out of the six films up to 45 Tesla, in order to extract information on the scattering times in the two types of bands. Indeed, the IRR40 sample, which has the highest $\rho_0$ and the lowest $T_c$, was expected not to show appreciable magnetoresistivity even at the highest magnetic field and thus it was not included in this analysis. In figure 2 the experimental magnetoresistivity data are presented. In the second to fourth panels, the series of neutron irradiated samples is presented. It can be seen that the IRR10 sample, which in practice can be considered as almost unirradiated, has a very large magnetoresistivity which reaches 100% for the B⊥$ab$ configuration. In the B∥$ab$ configuration the magnetoresistivity is smaller by a factor of nearly two. With increasing irradiation-induced disorder, the magnetoresistivity curves in the two configurations become closer and closer to each other, in the case of the IRR20 sample the B∥$ab$ curve is even slightly above the B⊥$ab$ curve, but we do not dare to speculate on the origin of this crossing as it is hardly outside the experimental uncertainty. The magnetoresistivity decreases and becomes more and more linear in $B^2$ with increasing irradiation, as expected when the scattering times are much smaller than the inverse cyclotron frequency. Its value is nearly 30% for the IRR20 sample and one order of magnitude smaller for the IRR30 sample. The thin sample TS, shown in the uppermost panel of figure 2, exhibits a magnetoresistivity which reaches 10% at 45 Tesla in the B⊥$ab$ configuration and only 3% in the B∥$ab$ configuration. Finally the carbon-doped sample has a 5% magnetoresistivity for both configurations and, again, the $B^2$ dependence characteristic of the most disordered samples. The magnitude of the magnetoresistivity curves in the two orientations is well described by the fitting curves calculated by the expression shown in section II and plotted as continuous lines in figure 2. On the other hand, the curvature of the experimental data is not well reproduced by the fits, especially for the less resistive samples. This can be partially due to the uncertainty on the theoretical values assumed for effective masses and charge densities; however, also the details of the shape of the Fermi surface possibly play a role, making the approximation of diagonal effective mass tensor inadequate.

The values of the scattering times ratio $b=\tau_\pi/\tau_\sigma$ obtained by the fits are reported in table III. We find *b=0.6*, *2.5* and *1.75* for IRR10, IRR20 and IRR30, respectively. This seems to indicate that for the weakly neutron irradiated sample IRR10 the π-bands are dirtier; with increasing irradiation, the σ-bands are progressively more affected by disorder than the π-bands; eventually, for the highly irradiated sample all the bands are equally disordered. The fit provides reasonable values for the parameter *b* also for the case of the other two samples. In the thin sample TS the value *b=0.8* also indicates dirtier π-bands, as in the less irradiated sample, suggesting that this is a characteristic peculiar of as-grown films. Finally, for the carbon doped sample we find *b=1.65*, which again means nearly equally disordered bands. The absolute values of the scattering times extracted from the fitting parameters *a* and *b* are also listed in table II for the five samples. Since the inverse cyclotron frequency at 40 Tesla is of the order of $10^{-14}$ s, these results are consistent with the $B^2$ dependence observed in the most resistive samples.

The in-plane residual resistivity can be calculated from the fitting parameters by equation (2) and compared with the experimental value. The calculated and experimental resistivities are reported in tables III and II, respectively. It can be seen that they are in reasonable agreement, except for the carbon doped sample. This can be due to the fact that in this case the measured resistivity is not the intrinsic one, but it is enhanced by the highly resistive amorphous areas at the grain boundaries. The general agreement between the calculated and experimental resistivities is a good check of the reliability of this magnetoresistivity analysis.

As another independent check of the results, the scattering times can be obtained from the temperature dependence of the resistivity, which reflects the contribution of different bands to the transport. In clean $MgB_2$ the room temperature derivative of the resistivity ρ'(300K) is expected to be very close to the value *$\rho_\pi'$=0.06 μΩ·cm/K* predicted for π-bands only[18,29]. Indeed, as σ–bands carriers are more strongly coupled to phonons, the temperature derivative of the part of the resistivity contributed by the σ-bands is steeper, so that eventually the σ-bands cease to contribute to transport at high temperatures, in a picture of independent parallel conduction channels. The room temperature derivative approaches the π-bands value also in the case of samples with dirty σ-bands, where σ-bands contribute even less to transport, and in the case of samples with equally dirty σ-and π-bands as well; indeed, room temperature transport in π-bands switches to the parallel σ channel only if this latter is sufficiently clean; oppositely, if both bands are dirty, the less steep temperature dependence of the π-bands makes again the π-bands the dominant conduction channel at room temperature. In table II the room temperature derivatives of the resistivity curves are reported; it can be seen that the value for the thin sample TS is larger than $\rho_\pi'$ and it decreases, approaching $\rho_\pi'$ for the irradiated samples. This behavior is compatible with the behavior of the scattering times extracted from the magnetoresistivity analysis; namely, it is consistent with a picture where the unirradiated samples have slightly dirtier π-bands, while in the increasingly irradiated samples the disorder affects both bands more or less equally. In the carbon doped sample, the extrinsic contribution to resistivity makes any speculation on bands contributions unreliable.

## V. Conclusions

In summary, we demonstrate that normal state magnetoresistivity measured in high magnetic fields both parallel and perpendicular to the *ab* planes is a useful tool to extract the relative scattering times in the different bands in magnesium diboride oriented thin films. In particular we apply this method to study the effect of controlled disorder introduced by neutron irradiation and carbon doping. The result of our analysis is that the undoped unirradiated thin films have cleaner σ-bands; upon irradiation both bands become increasingly dirtier; eventually highly irradiated samples and carbon doped samples have comparable scattering times in the two types of bands. Room temperature derivative of the resistivity curves and residual resistivity values are consistent with this description. Furthermore, in a forthcoming work, these results will be useful to understand the behavior of upper critical fields in these samples as a function of disorder.

**Figure captions**

**Table I:** Effective masses for both crystal directions, as well as charge densities per unit cell, for all four bands of $MgB_2$ crossing the Fermi surface.

**Table II:** List of the samples with neutron fluences in $cm^{-2}$ (for the irradiated samples only), critical temperatures $T_c$ (defined at 50% of normal state resistivity) and transition widths $\Delta T_c$ in K, experimental in plane residual resistivities $\rho_{in\text{-}plane}^{(exp)}$ in $\mu\Omega\cdot cm$ measured at 42K, residual resistivity ratios RRR and temperature derivatives of resistivity curves $\rho'$ at 300K in $\mu\Omega\cdot cm/K$.

**Table III:** List of the samples with the fitting parameters $b=\tau_\pi/\tau_\sigma$ and $a=|\mu_{(c)\pi l}|$ used to describe the experimental curves shown in figure 2 by equations (1), as well as the in-plane residual resistivity $\rho_{in\text{-}plane}^{(calc)}$ calculated by equation (2) and the absolute values of the scattering times in $\pi$- and $\sigma$-bands calculated as $|\mu_{(dir)i}|\cdot m_{eff(dir)i}/e$ in seconds.

**Figure 1:** Calculated magnetoresistivity curves for the $B\perp ab$ and $B||ab$ configurations for three different values of the parameter $b=\tau_\pi/\tau_\sigma$ equal to *0.5*, *1.65* and *5.5* from the top panel to the bottom one and for values of the parameters $a=|\mu_{\pi l}^{(c)}|$ equal to *0.0059*, *0.0099* and *0.0125* $m^2\cdot V^{-1}\cdot s^{-1}$ respectively, such that the in-plane resistivity turns out to be $\rho=5$ $\mu\Omega\cdot cm$ in all cases. In other

words, from top to bottom the σ-bands become increasingly dirty with respect to π-bands are shown.

**Figure 2:** Experimental magnetoresistivity curves for the $B \perp ab$ (open squares symbols) and $B||ab$ (filled round symbols) configurations; also curves calculated with the fitting parameters listed in table III are shown as continuous lines.

| Band | $m_{eff}/m_0$ | | $n/V$ |
|---|---|---|---|
| | // ab planes | // c axis | |
| σ1 | 0.167 | 33.21 | 0.051 |
| σ2 | 0.377 | 106.01 | 0.100 |
| π1 | 0.439 | 0.314 | 0.265 |
| π2 | 0.538 | 0.153 | 0.114 |

| Sample | Neutron fluence $(cm^{-2})$ | $T_c$ (K) | $\Delta T_c$ (K) | $\rho_{in\text{-}plane}^{(exp)}$ $(\mu\Omega \cdot cm)$ | RRR | $\rho'(300K)$ $(\mu\Omega \cdot cm/K)$ |
|---|---|---|---|---|---|---|
| TS | -- | 39.4 | 0.6 | 7.0 | 3.9 | 0.093 |
| IRR10 | $6.4 \cdot 10^{15}$ | 41.05 | 0.1 | 1.0 | 9.8 | 0.053 |
| IRR20 | $6.5 \cdot 10^{16}$ | 40.7 | 0.2 | 2.3 | 4.5 | 0.045 |
| IRR30 | $7.7 \cdot 10^{17}$ | 36.1 | 0.4 | 16 | 1.6 | 0.062 |
| IRR40 | $9.5 \cdot 10^{18}$ | 17 | 4.0 | 69 | 1.1 | 0.056 |
| C-dop | -- | 38.5 | 1.0 | 20 | 2.3 | 0.110 |

| Sample | b | a ($m^2 \cdot V^{-1} \cdot s^{-1}$) | $\rho_{in-plane}^{(calc)}$ ($\mu\Omega \cdot cm$) | $\tau_\pi$ (s) | $\tau_\sigma$ (s) |
|---|---|---|---|---|---|
| TS | 0.80 | 0.0070 | 5.4 | $1.3 \cdot 10^{-14}$ | $1.6 \cdot 10^{-14}$ |
| IRR10 | 0.60 | 0.0270 | 1.2 | $4.8 \cdot 10^{-14}$ | $8.0 \cdot 10^{-14}$ |
| IRR20 | 2.50 | 0.0205 | 2.7 | $3.7 \cdot 10^{-14}$ | $1.5 \cdot 10^{-14}$ |
| IRR30 | 1.75 | 0.0055 | 9.2 | $9.8 \cdot 10^{-15}$ | $5.6 \cdot 10^{-15}$ |
| C-dop | 1.65 | 0.0070 | 7.0 | $1.3 \cdot 10^{-14}$ | $7.7 \cdot 10^{-15}$ |

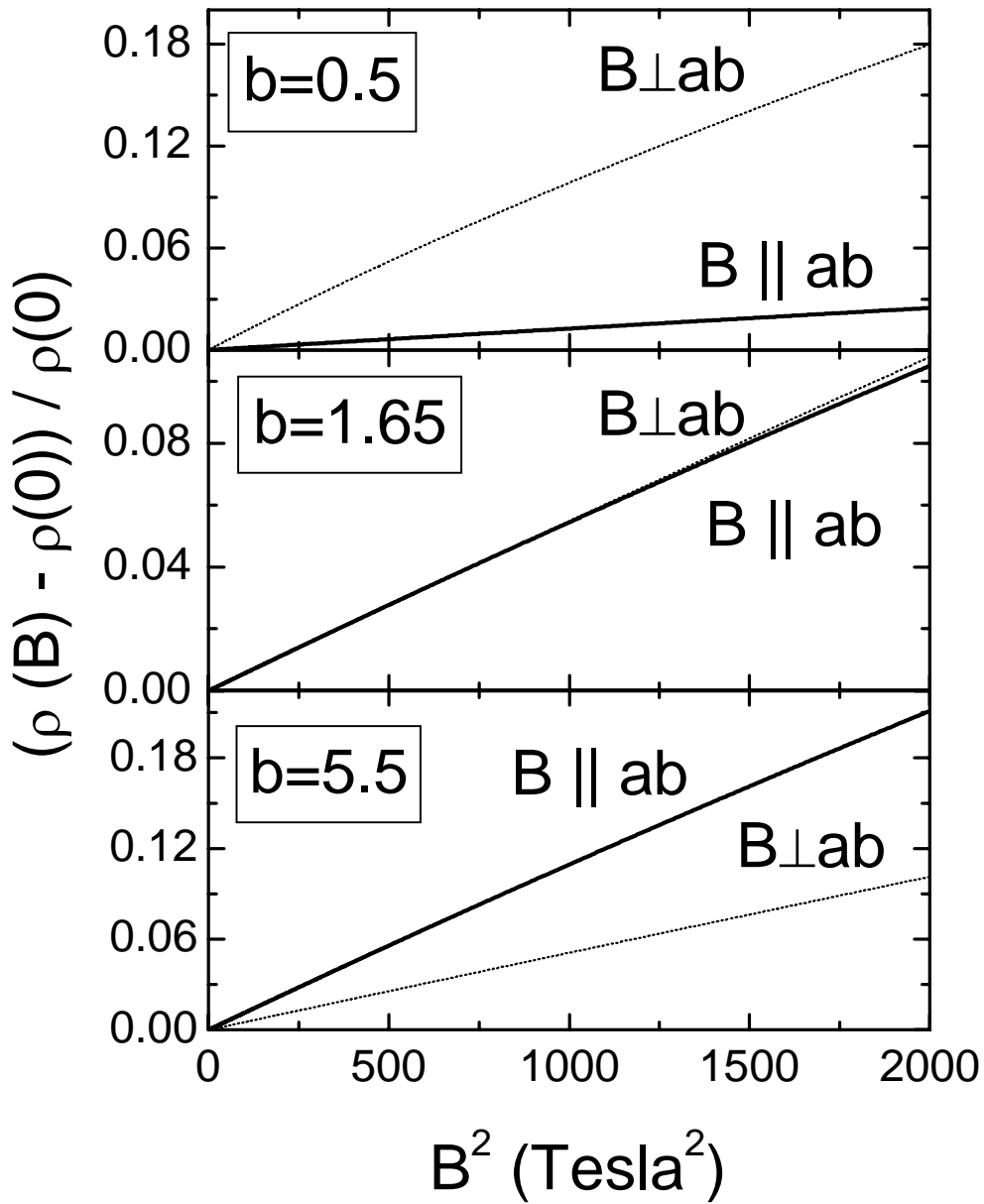

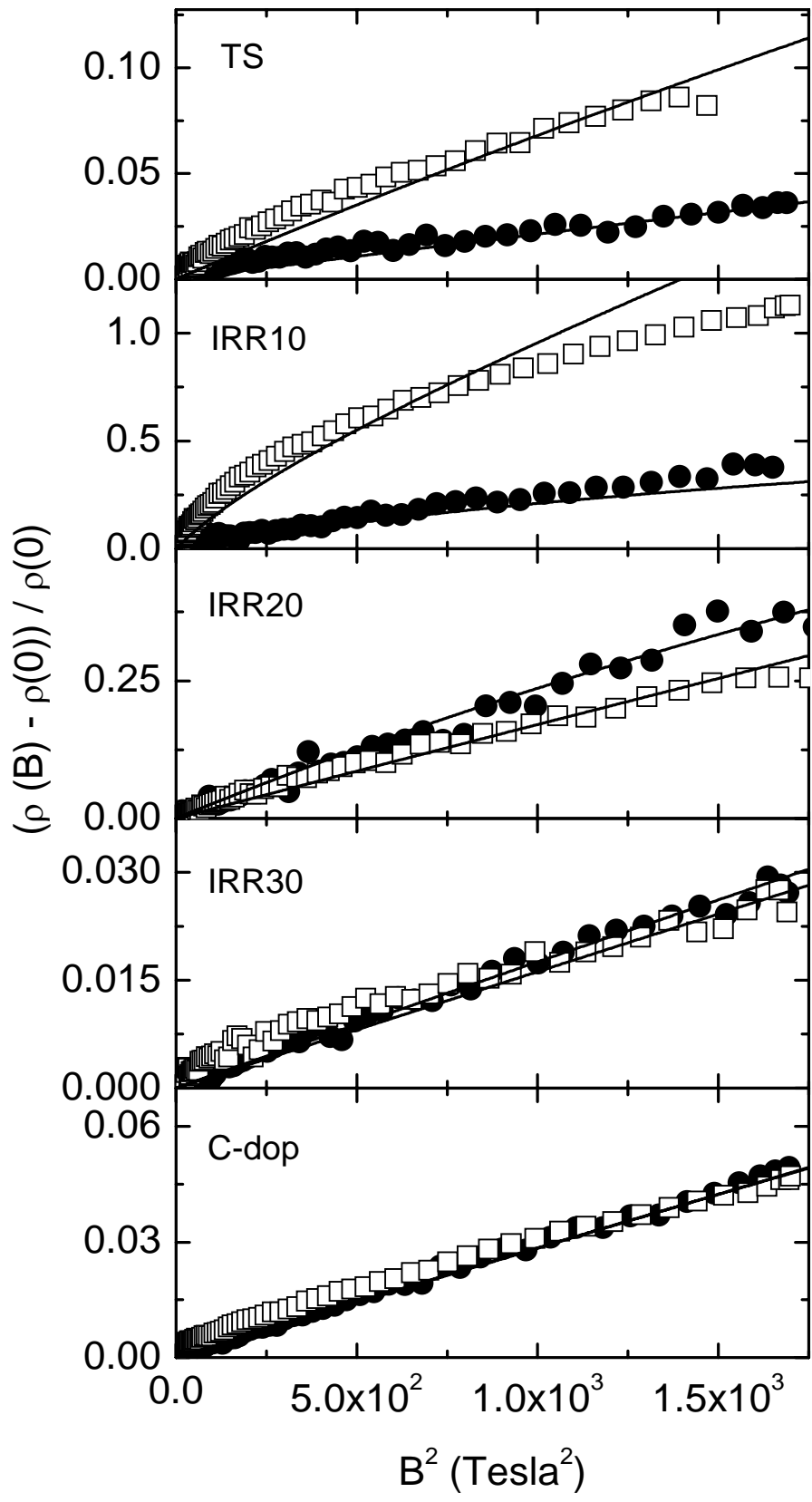